\documentclass[12pt]{article}
\def\decpk{$\Theta^+ \rightarrow K^0 p$}
\def\decnk{$\Theta^+ \rightarrow K^+ n$}
\def\decpkg{$\Theta^+ \rightarrow K^0 p \gamma$}
\def\decnkg{$\Theta^+ \rightarrow K^+ n \gamma$}
\def\widpk{$\Gamma(\Theta^+ \rightarrow K^0 p)$}
\def\widnk{$\Gamma(\Theta^+ \rightarrow K^+ n)$}
\def\widNK{$\Gamma(\Theta^+ \rightarrow KN)$}
\def\widpkg{$\Gamma(\Theta^+ \rightarrow K^0 p \gamma)$}
\def\widnkg{$\Gamma(\Theta^+ \rightarrow K^+ n \gamma)$}
\def\widNKg{$\Gamma(\Theta^+ \rightarrow KN \gamma)$}
\def\ks_pipi{$K^0_S \rightarrow \pi^+\pi^-$}

\def\ypiz_2gg{$\pi^0 \rightarrow \gamma\gamma$}

\begin {document}

\title
{Experimental search for radiative decays of the pentaquark baryon
$\Theta^+(1540)$}

\author{
DIANA Collaboration\\
V.V. Barmin$^a$, 
A.E. Asratyan$^a$,
V.S. Borisov$^a$, 
C. Curceanu$^b$, \\
G.V. Davidenko$^a$, 
A.G. Dolgolenko$^{a,}$\thanks{Corresponding author. E-mail address:
dolgolenko@itep.ru.},
C. Guaraldo$^b$, 
M.A. Kubantsev$^a$, \\
I.F. Larin$^a$, 
V.A. Matveev$^a$, 
V.A. Shebanov$^a$, 
N.N. Shishov$^a$, \\
L.I. Sokolov$^a$,
G.K. Tumanov$^a$,
and V.S. Verebryusov$^a$ \\
\normalsize {$^a$ \it Institute of Theoretical and Experimental Physics,
Moscow 117259, Russia}\\
\normalsize {$^b$ \it Laboratori Nazionali di Frascati dell' INFN,
C.P. 13-I-00044 Frascati, Italy}
}                                          
\date {\today}
\maketitle

\begin{abstract}
The data on the reactions 
$K^+\mathrm{Xe} \rightarrow K^0 \gamma X$  and
$K^+\mathrm{Xe} \rightarrow K^+ \gamma X$,
obtained with the bubble chamber DIANA, have been analyzed for 
possible radiative decays of the $\Theta^+(1540)$ baryon:
\decpkg\ and \decnkg. No signals have been observed, and we derive the
upper limits \widpkg/\widpk\ $< 0.032$ and \widnkg/\widnk\ $< 0.041$
which, using our previous measurement of \widNK = $(0.39\pm0.10)$ MeV,
translate to \widpkg\ $< 8$ keV and \widnkg\ $< 11$ keV
at 90\% confidence level. We have also measured the
cross sections of $K^+$-induced reactions involving emission of a
neutral pion:
$\sigma(K^+n \rightarrow K^0 p \pi^0 ) = (68\pm18)$ $\mu$b and
$\sigma(K^+N \rightarrow K^+ N \pi^0 ) = (30\pm8)$ $\mu$b 
for incident $K^+$ momentum of 640 MeV.
\end{abstract}

\newpage

     Possible existence of multiquark hadrons, and pentaquark baryons
in particular, has been discussed for many years \cite{forerunners}. 
Fairly definite predictions for the antidecuplet of light pentaquark
baryons with spin--parity ${1/2}^+$ have been formulated by Diakonov, 
Petrov, and Polyakov in the framework of the chiral quark--soliton model 
\cite{DPP}. In particular, they predicted $m \simeq 1530$ MeV and 
$\Gamma < 15$ MeV for the mass and width of the explicitly exotic baryon 
with $S = +1$ and $I = 0$, the $\Theta^+(uudd\bar{s})$ that should decay 
to the $nK^+$ and $pK^0$ final states. 
Narrow peaks near 1540 MeV in the $nK^+$ and $pK^0$ mass spectra were 
initially detected in low-energy photoproduction by LEPS \cite{Nakano-old} 
and in the charge-exchange reaction $K^+n \rightarrow pK^0$ by DIANA
\cite{DIANA-old}. 
Other searches for the $\Theta^+$ baryon in different reactions and
experimental conditions yielded positive evidence as well as null
results casting doubt on its existence, see the review papers 
\cite{Burkert} and \cite{Danilov-Mizuk}. However, both LEPS and DIANA 
were able to confirm their initial observations 
\cite{Nakano,DIANA,DIANA-anew}. 

     Using the unique properties of the charge-exchange reaction 
that forms $\Theta^+$ baryons in the $s$-channel, DIANA reported a 
direct measurement of the $\Theta^+$ intrinsic width: 
$\Gamma = 0.39\pm0.10$ MeV \cite{DIANA,DIANA-anew}. 
The extreme smallness of the $\Theta^+$ width, as compared with
well known baryons, indicates a strong
dynamical suppression of the ``fall-apart" mechanism of $\Theta^+$ 
decays to $pK^0$ and $nK^+$. This does not necessarily imply that 
radiative decays to $\gamma pK^0$ and $\gamma nK^+$ should be equally
suppressed, so that the relative probability of radiative decay may
prove to be higher for the $\Theta^+$ baryon than for ordinary hadrons.
Indeed, partial widths of the radiative decays \decpkg\ and \decnkg\ have
been computed in \cite{Hong} as \widnkg\ = 34--41 keV and 
\widpkg\ $= 0.25\times$\widnkg, 
so that \widnkg/\widnk\ $\simeq 0.15$ if \widnk\ = $0.5\times$\widNK\ 
and the measured value of \widNK\ \cite{DIANA-anew} is substituted.
On the other hand, the $\Theta^+$ radiative widths predicted in
\cite{He,Ioffe} using different assumptions
are smaller than in \cite{Hong} by nearly two orders
of magnitude\footnote{An important 
conclusion reached in \cite{Ioffe} was 
that the parity of the $\Theta^+$ 
baryon can be determined from the
shape of the $\gamma$ energy spectrum.}.
Anyway, the simple arguments brought above and the prediction 
\cite{Hong} for \widNKg\ provide adequate motivation for a search for 
radiative decays \decpkg\ and \decnkg\ in the DIANA experiment. The 
results of the search are reported in this paper.

     The non-magnetic bubble chamber DIANA \cite{chamber} 
was filled with liquid Xenon and exposed to a separated $K^+$ beam with 
incident momentum of 850 MeV at the proton synchrotron at ITEP, Moscow.
The fiducial volume of the chamber was $700\times700\times1400$ mm$^3$,
and the density and radiation length of the fill were 2.2 g/cm$^3$ and
3.2 cm, respectively. The total number of incident $K^+$ mesons was
$\sim 10^6$, and the data were recorded on $\sim 4*10^5$ stereo-frames.
In the fiducial volume of the bubble chamber, $K^+$ momentum varies 
from $\sim 730$ MeV for entering kaons to zero for those that range
out through ionization. Throughout this momentum interval, all collisions
and decays of incident $K^+$ mesons are efficiently detected. The $K^+$
momentum at interaction point is determined from the spatial distance
between the detected vertex and the center of the observed maximum due
to decays of stopping $K^+$ mesons. Secondary protons and charged pions
are identified by ionization, charged kaons --- by ionization and decays,
$K^0_S$ mesons --- by the ranges and emission angles of decay pions, and 
$\gamma$-quanta --- by conversion to an $e^+e^-$ pair. Charged particles
are momentum-analyzed by range, and $\gamma$-quanta --- by the summed
length of the electron--positron shower. The apparatus efficiency for
$\gamma$-quanta with $E_\gamma > 30$ MeV is close to 100\%. Further
details on the experimental procedure may be found in
\cite{exp-procedure-1,exp-procedure-2,DIANA}.

     Radiative decays of the $\Theta^+$ baryon were searched for among the 
collisions with one or more detected $\gamma$-quanta in the final state:
$K^+\mathrm{Xe} \rightarrow K^+ (m\gamma) X$ and
$K^+\mathrm{Xe} \rightarrow K^0_S (m\gamma) X$, 
\ks_pipi, where $m \geq 1$.
We scanned for and then analyzed the $K^+$Xe collisions that showed
either a secondary $K^+$ or a Vee from \ks_pipi, plus at least one 
$e^+e^-$ pair from $\gamma$ conversion that pointed back to the primary
vertex. The collisions involving \ks_pipi\ were selected in the full data
sample, and those with a secondary $K^+$ --- in nearly 30\% of the data.
The events were selected in a double scan of the film, and a portion of 
the film was scanned three times for better understanding of the scanning
efficiency. All selected events were then carefully measured and 
reconstructed in space using specially designed stereo-projectors. In 
the interval $350 < p(K^+) < 600$ MeV of the incident $K^+$ momentum, we 
found 3 events of the reaction
$K^+\mathrm{Xe} \rightarrow K^0_S \gamma X$, \ks_pipi\ 
and 11 events of the reaction
$K^+\mathrm{Xe} \rightarrow K^+ \gamma X$ 
with $E_\gamma > 30$ MeV. We also found 20 events of the reaction
$K^+\mathrm{Xe} \rightarrow K^0_S X$ with \ks_pipi$\gamma$.

     The dominant source of background for the radiative decays
\decpkg\ and \decnkg\ are the radiative non-resonant transitions
$K^+n \rightarrow K^0 p \gamma$,
$K^+n \rightarrow K^+ n \gamma$, 
and $K^+p \rightarrow K^+ p \gamma$.
We estimate the probabilities of these transitions relative to 
corresponding binary reactions
$K^+n \rightarrow K^0 p$,
$K^+n \rightarrow K^+ n$,
and $K^+p \rightarrow K^+ p$
using the formalism developed in \cite{Pomeranchuk} 
(see also \cite{Ioffe}). The cross-section ratios 
$\sigma(K^+n \rightarrow K^0 p \gamma) / \sigma(K^+n \rightarrow K^0 p)$,
$\sigma(K^+n \rightarrow K^+ n \gamma) / \sigma(K^+n \rightarrow K^+ n)$,
and 
$\sigma(K^+p \rightarrow K^+ p \gamma) / \sigma(K^+p \rightarrow K^+ p)$
increase with incident momentum $p(K^+)$ by a factor $\sim 2$ between
the boundary values $p(K^+) = 350$ and 600 MeV. For our experimental
conditions, the yields of the radiative reactions
$K^+n \rightarrow K^0 p \gamma$,
$K^+n \rightarrow K^+ n \gamma$, 
and $K^+p \rightarrow K^+ p \gamma$
relative to 
$K^+n \rightarrow K^0 p$,
$K^+n \rightarrow K^+ n$,
and $K^+p \rightarrow K^+ p$
are computed as $3.8\times10^{-4}$, $5.6\times10^{-4}$, and 
$7.6\times10^{-4}$, respectively. 

     The experimental determination of the probabilities of radiative
processes
$K^+\mathrm{Xe} \rightarrow K^0_S \gamma X$
and $K^+\mathrm{Xe} \rightarrow K^+ \gamma X$ 
relative to the charge-exchange and elastic reactions
$K^+\mathrm{Xe} \rightarrow K^0_S X$
and $K^+\mathrm{Xe} \rightarrow K^+ X$
relies on the scanning information for the total numbers of 
events with detected \ks_pipi\ decays or secondary $K^+$ mesons. In 
the interval $350 < p(K^+) < 600$ MeV of the incident $K^+$ momentum, 
the scan found $12750\pm750$ events of the charge-exchange reaction
with \ks_pipi. With three events of the radiative reaction
$K^+\mathrm{Xe} \rightarrow K^0_S \gamma X$ 
detected with an efficiency 
$\epsilon(K^0_S\gamma X) = 0.65\pm0.06$, 
we obtain the relative
probability
\begin{displaymath}
W(K^+\mathrm{Xe} \rightarrow K^0_S \gamma X) /
W(K^+\mathrm{Xe} \rightarrow K^0_S X) = (3.6\pm2.1)\times10^{-4}
\end{displaymath}
which agrees with the theoretical prediction. 
When extracting the total number of events of the elastic reaction
$K^+\mathrm{Xe} \rightarrow K^+ X$
from the scanning information, the events due to coherent scattering
$K^+$Xe $\rightarrow K^+$Xe are rejected by the cut $T' < 0.9T$
where $T$ and $T'$ are kinetic energies of the incident and secondary
kaons. The number of events of the elastic reaction corrected for the
cut $T' < 0.9T$ is estimated as $29500\pm2100$ events for
$350 < p(K^+) < 600$ MeV. With 11 events of the corresponding radiative 
reaction
$K^+\mathrm{Xe} \rightarrow K^+ \gamma X$ 
detected with an efficiency 
$\epsilon(K^+\gamma X) = 0.75\pm0.05$, 
we obtain the relative probability
\begin{displaymath}
W(K^+\mathrm{Xe} \rightarrow K^+ \gamma X) /
W(K^+\mathrm{Xe} \rightarrow K^+ X) = (5.0\pm1.7)\times 10^{-4}.
\end{displaymath}
The latter ratio is in fact the mean value of relative probabilities for
the proton and neutron, because the numbers of elastic $K^+p$ and $K^+n$
collisions are expected to be virtually equal (the proportion between
protons and neutrons in the Xenon nucleus proves to be very close to
the inverse ratio between the $K^+p$ and $K^+n$ elastic cross sections
\cite{Bowen,Casadei}.) Again, the measured value agrees with the 
theoretical prediction of $6.6\times10^{-4}$ (the mean value for the
proton and neutron, see above). The number of detected charge-exchange 
events with \ks_pipi$\gamma$ ($20\pm5$) agrees with the 
results of a simulation\footnote{The 
simulation reproduces the experimental
conditions and selections, relies on the 
total number of detected events with 
\ks_pipi\ decays for normalization, and 
accounts for the branching fraction of 
the decay \ks_pipi$\gamma$ \cite{PDG}.}
that predicts 23 events of this type. We may conclude that the detected
$K^+\mathrm{Xe} \rightarrow K^0_S \gamma X$ 
and $K^+\mathrm{Xe} \rightarrow K^+ \gamma X$
events are consistent with being entirely due to the non-resonant 
radiative transitions
$K^+n \rightarrow K^0 p \gamma$,
$K^+n \rightarrow K^+ n \gamma$, 
and $K^+p \rightarrow K^+ p \gamma$.

     In estimating the probabilities of radiative decays \decpkg\ and
\decnkg, we rely on our previous observation of $71.7\pm13.2$ decays 
\decpk, \ks_pipi\ in the $K^+$ momentum range of 445--525 MeV 
\cite{DIANA-anew}. Correcting for the scanning efficiency of 0.83
for the reaction
$K^+\mathrm{Xe} \rightarrow K^0_S X$
and for unmeasurable events whose fraction
reaches $0.40\pm0.07$, we obtain a ``reference" signal of \\
\hspace*{1.2in} $N_0$(\decpk, \ks_pipi) = $(145\pm27)$ events \\
for $445 < p(K^+) < 525$ MeV.
To derive the ``reference" number of \decnk\ decays, we assume equal
branching fractions for \decpk\ and \decnk, correct for the probability
of $K^0 \rightarrow \pi^+\pi^-$ and for the difference of detection
efficiencies for events with secondary $K^+$ and \ks_pipi, and take
into account that the final state $K^+ p \gamma$ has been analyzed
for only 30\% of the data. Thereby, for $445 < p(K^+) < 525$ MeV
we obtain \\
\hspace*{1.2in} $N_0$(\decnk) = $(127\pm23)$ events.

     Of the three detected events of the reaction
$K^+\mathrm{Xe} \rightarrow K^0_S \gamma X$,
only one lies within the interval 445--525 MeV of incident $K^+$
momentum. The background from the the non-resonant radiative
transition 
$K^+n \rightarrow K^0 p \gamma$ 
is estimated as 1.0 events. The final state of the observed event is 
$K^0_S p \gamma$ as expected for \decpkg, but the effective mass of this 
three-body system ( $m_3 = (1515\pm12)$ MeV) is below 
$m(\Theta^+) = (1538\pm2)$ MeV \cite{DIANA-anew} by almost 2$\sigma$. 
Therefore, we disregard this event and conclude that the number of 
detected \decpk\ decays is less than 2.3 at 90\% confidence level. 
Dividing by the detection efficiency $\epsilon(K^0_S\gamma X)$ and by the 
reference signal $N_0$(\decpk, \ks_pipi) as quoted above, we obtain \\
\hspace*{0.8in} \widpkg/\widpk\ $< 0.032$ at 90\% confidence level. \\
Of the 11 detected events of the reaction
$K^+\mathrm{Xe} \rightarrow K^+ \gamma X$,
three fall within the incident momentum interval of
$445 < p(K^+) < 525$ MeV (two events are at $350 < p(K^+) < 445$ MeV 
and six --- at $525 < p(K^+) < 600$ MeV).
The background from non-resonant transitions
$K^+N \rightarrow K^+ N \gamma$ is estimated as 4.7 events, so the 
number of detected \decnk\ decays is less than 3.0 at 90\% confidence 
level \cite{Feldman-Cousins}. Dividing
by the detection efficiency $\epsilon(K^+\gamma X)$ and by the 
corresponding reference signal $N_0$(\decnk), we obtain \\
\hspace*{0.8in} \widnkg/\widnk\ $< 0.041$ at 90\% confidence level. \\
Substituting \widpk\ = \widnk\ = $(0.19\pm0.05)$ MeV as measured in
\cite{DIANA-anew}, we are able to restrict the absolute partial widths
of $\Theta^+$ radiative decays: \widpkg\ $< 8$ keV and 
\widnkg\ $< 11$ keV at 90\% confidence level. These measurements are
in obvious disagreement with the ``optimistic" predictions 
\widpkg\ = 8--10 keV and \widnkg\ = 34--41 keV \cite{Hong}. On the
other hand, the sensitivity of our experiment is insufficient for a 
substantiative comparison with lower predictions \cite{He,Ioffe}.

     Apart from the radiative reactions
$K^+\mathrm{Xe} \rightarrow K^0_S \gamma X$
and $K^+\mathrm{Xe} \rightarrow K^+ \gamma X$,
we have also searched for the inelastic collisions involving formation
of neutral pions. In the interval $540 < p(K^+) < 660$ MeV of the
incident $K^+$ momentum, 22 events of the reaction
$K^+\mathrm{Xe} \rightarrow K^0_S \pi^0 X$, \ks_pipi
and 27 events of the reaction 
$K^+\mathrm{Xe} \rightarrow K^+ \pi^0 X$
have been found. The distribution of these inelasic events in 
incident $K^+$ momentum steeply increases with $p(K^+)$ from the 
threshold of $p(K^+) \simeq 520$ MeV to the maximum measured value of 
$p(K^+) = 660$ MeV. The energy spectrum of detected $\pi^0$ mesons
agrees with that for the simulated reaction
$K^+N \rightarrow KN\pi^0$ 
on a bound nucleon \cite{Zorn}. Selecting the beam-momentum interval 
$600 < p(K^+) < 660$ MeV with $\langle p(K^+) \rangle \simeq 640$ MeV
and taking into account the total numbers of events due to the 
charge-exchange and elastic reactions
$K^+\mathrm{Xe} \rightarrow K^0_S X$
and $K^+\mathrm{Xe} \rightarrow K^+ X$ 
(not quoted), for the reactions with $\pi^0$ emission on the Xenon 
nucleus we obtain
$W(K^+\mathrm{Xe} \rightarrow K^0_S \pi^0 X) /
W(K^+\mathrm{Xe} \rightarrow K^0_S X) = (0.90\pm0.23)$\%
and
$W(K^+\mathrm{Xe} \rightarrow K^+ \pi^0 X) /
W(K^+\mathrm{Xe} \rightarrow K^+ X) = (0.31\pm0.07)$\%
at $p(K^+) = 640$ MeV.
These relative yields have been corrected for the scanning efficiency 
for the reactions of $\pi^0$ emission and for $\pi^0$ absorption in
the Xenon nucleus \cite{Hornbostel}. Substituting the measured cross
sections of the binary reactions at $p(K^+) = 640$ MeV as
$\sigma(K^+n \rightarrow K^0 p) = (7.5\pm0.5)$ mb,
$\sigma(K^+n \rightarrow K^+ n) = (8\pm1)$ mb, 
and $\sigma(K^+p \rightarrow K^+ p) = (12.5\pm1.0)$ mb
\cite{Slater,Bowen,Casadei} and taking into account that the Xe nucleus
consists of $Z = 54$ protons and $A - Z = 77$ neutrons, we obtain \\
\hspace*{0.9in}$\sigma(K^+n \rightarrow K^0 p \pi^0 ) =(68\pm18)$ $\mu$b,\\
\hspace*{0.9in}$\sigma(K^+N \rightarrow K^+ N \pi^0 ) =(30\pm8)$ $\mu$b, \\
where the latter refers to the cross section averaged over constituent
nucleons of the Xe nucleus.
The cross sections of these reactions for $p(K^+) \simeq 640$ MeV
were previously measured in a single experiment that used a deuterium 
bubble chamber \cite{Giacomelli}:
$\sigma(K^+n \rightarrow K^0 p \pi^0 ) = (80\pm20)$ $\mu$b,
$\sigma(K^+d \rightarrow K^+ \pi^0 p n ) = (0\pm20)$ $\mu$b.
The former value is in good agreement with our measurement.

     In summary, we have analyzed the data on the reactions
$K^+\mathrm{Xe} \rightarrow K^0 \gamma X$  and
$K^+\mathrm{Xe} \rightarrow K^+ \gamma X$,
obtained with the bubble chamber DIANA, for 
possible radiative decays of the $\Theta^+(1540)$ baryon:
\decpkg\ and \decnkg. No signals have been observed, and we derive the
upper limits \widpkg/\widpk\ $< 0.032$ and \widnkg/\widnk\ $< 0.041$
which, upon using our previous measurement of \widNK, translate to 
\widpkg\ $< 8$ keV and \widnkg\ $< 11$ keV at 90\% confidence level.
We have also measured the cross sections of $K^+$-induced reactions 
involving emission of a neutral pion:
$\sigma(K^+n \rightarrow K^0 p \pi^0 ) = (68\pm18)$ $\mu$b and
$\sigma(K^+N \rightarrow K^+ N \pi^0 ) = (30\pm8)$ $\mu$b 
for incident $K^+$ momentum of 640 MeV.

     It is a pleasure to thank L.~N. Bogdanova, B.~L. Ioffe, 
Yu.~S. Kalashnikova, A.~V. Samsonov, and I.~I. Strakovsky for useful 
discussions and comments. This work is supported by the Russian 
Foundation for Basic Research (grant 07-02-00684).

\end{document}